\documentstyle[preprint,floats,prd,aps]{revtex}
\input epsf.tex
\def\DESepsf(#1 width #2){\epsfxsize=#2 \epsfbox{#1}}
\begin{document}
\preprint{\vbox {\hbox{OCHA-PP-69}}}
\draft
\title{Momentum Dependent Photon-Vector Meson Coupling and Parity
Violating effects in $B\to X_s\ell^+\ell^-$
\footnote{Work supported by the Japanese Society for the Promotion of
Science.}} %
\author{Mohammad R. Ahmady}\address{Department of Physics \\
Ochanomizu University \\
1-1 Otsuka 2, Bunkyo-ku,Tokyo 112, Japan}

\date{October 1995}
\maketitle
\begin{abstract}
We examine the leptonic forward-backward and
polarization (left-right) asymmetries in the
dileptonic $B\to X_s\ell^+\ell^-$ decay when the momentum dependence of
$\psi$ and $\psi'-\gamma$ conversion strength is taken into account.  The
results indicate only a small shift in the asymmetry distributions.
\end{abstract}
In a recent work \cite{prd}, we have shown that, when the momentum
dependence of photon-vector meson coupling is taken into account, the
"resonance to nonresonance" interference in the dileptonic invariant mass
distribution of $B\to X_s\ell^+\ell^-$ is substantially reduced.  This
momentum dependence, which is necessary to explain the data on $\psi$
leptonic width and photoproduction simultaneously \cite{terasaki}, is
also believed to suppress the long-distance (LD) contributions to $b\to
s\gamma$ decay \cite{dht,eims}.  The significance of this investigation
is due to the fact that the CKM favored intermediate $\psi (NS)$ vector
mesons contribute to $b\to s$ transitions by conversion to real($b\to
s\gamma$ decay) or virtual($b\to s\ell^+\ell^-$ decay) photons [5-8].
These
resonance contributions, in fact, dominate the total dileptonic decay
rates.  Therefore, to probe the contributing short-distance (SD)
operators for, among other things, signals of "new physics", one has to
take a careful account of the LD interference.

In this letter, we examine the effect of the momentum dependence of
$\psi$ and $\psi'-\gamma$ conversion strength on the parity-violating
aspects of the dileptonic rare B-decays.  The lepton pair
forward-backward asymmetry \cite{amm} and polarization
asymmetry distributions \cite{hewett} have been proposed as possible venues
to probe the SD contributions.  Our results indicate that, unlike the
invariant mass spectrum, these asymmetry distributions are not altered
significantly due to a momentum dependent photon-vector meson coupling.

We start with the low energy effective Lagrangian for $b\to
s\ell^+\ell^-$:
\begin{equation}
\displaystyle
L_{eff} =\frac {G_F}{\sqrt 2} \left ( \frac {\alpha }{4 \pi s_W^2}
\right ) V^*_{ts}V_{tb}
(A\bar s L_\mu b \bar \ell L^\mu \ell +B \bar s L_\mu b \bar\ell
R^\mu \ell +2m_b C \bar s T_\mu b \bar\ell \gamma^\mu\ell ),
\end{equation}
where
$$
L_\mu =\gamma_\mu (1-\gamma_5), \quad
R_\mu =\gamma_\mu (1+\gamma_5),
$$
and
$$
T_\mu = -i\sigma_{\mu \nu} (1+\gamma_5) q^\nu /q^2 .
$$
$V_{ij}$ are the Cabibbo-Kobayashi-Maskawa matrix elements,
$s_W^2=sin^2\theta_W\approx 0.23$ ($\theta_W$ is the weak angle), $G_F$
is the Fermi constant and $q$ is the total momentum of the final
$\ell^+\ell^-$ pair.

The SD parts of $A$ and $B$, denoted by $A^{\rm SD}$ and $B^{\rm SD}$, arise
from W box diagrams and penguin diagrams with Z gauge boson and photon
coupled to $\ell^+\ell^-$ pair.  For $m_t=180 GeV$, $m_b=4.5 GeV$ and
$\Lambda_{QCD}=100 MeV$ we obtain \cite {see}
\begin{equation}
\begin{array}{l}
A^{\rm SD}=2.020 ,\\
B^{\rm SD}=-0.173 ,\\
C=-0.146 .
\end{array}
\end{equation}

The LD part of $A$ and $B$ coefficients receive contributions from
charm quark loop ($c\bar c$ continuum), and $\psi$ and $\psi'$ resonances.
\begin{equation}
A^{\rm LD}=B^{\rm LD}=-s_W^2\left (3C_1(m_b)+C_2(m_b)\right )(\tau^{\rm
cont}+\tau^{\rm res})
\end{equation}
The combination of the Wilson coefficients in (3) is assigned a value
$$
\vert 3C_1(m_b)+C_2(m_b)\vert =0.72
$$
which fits the data on the semi-inclusive $B\to X_s\psi$ \cite{dht}. The
$c\bar
c$ continuum contribution is obtained from the electromagnetic penguin
diagrams \cite {gsw} \begin{equation}
\displaystyle
\tau^{\rm cont}=g\left
(\frac{m_c}{m_b}\; ,\; z\right )
\end{equation}
where $z=q^2/m_b^2$ and
\begin{equation}
\displaystyle
g(y\; ,\; z)= \left \{ \begin{array}{ll}
\displaystyle
-\left [ \frac{4}{9}\ln
(y^2)-\frac{8}{27}-\frac{16}{9}\frac{y^2}{z}+\frac{2}{9}\sqrt{1-\frac{4y^2}{z}}
\left (2+\frac{4y^2}{z}\right )\left (\ln \frac{
\vert 1+\sqrt{1-\frac{4y^2}{z}}\vert}{\vert 1-\sqrt{1-\frac{4y^2}{z}}\vert}
+i\pi\right )\right ] & \;  z\ge 4y^2 \\
\displaystyle
-\left [ \frac{4}{9}\ln
(y^2)-\frac{8}{27}-\frac{16}{9}\frac{y^2}{z}+\frac{4}{9}\sqrt{\frac{4y^2}{z}-1}
\left (2+\frac{4y^2}{z}\right ) arctan
\frac{1}{\sqrt{\frac{4y^2}{z}-1}}\right ]  & \;  z\le 4y^2

\end{array}
\right.
\end{equation}

The resonance contributions from $\psi$ and $\psi'$
can be incorporated by using a Breit-Wigner form for the resonance
propagator \cite{lms,dtp}:
\begin{equation}
\tau^{\rm res}=\frac{16\pi^2}{9}\left (\frac{f^2_\psi
(q^2)/m^2_\psi}{m^2_\psi-q^2-im_\psi\Gamma_\psi}+\; (\psi\to\psi')\right
)e^{i\phi}
\end{equation}
The relative phase $\phi$ that determines the sign between $\tau^{\rm
cont}$ and $\tau^{\rm res}$ is chosen to be zero due to unitarity
constraint \cite {ot}.

{}From (6) we observe that $\tau^{\rm res}$ depends
quadratically on $f_V(q^2) (V=\psi\;,\;\psi' \;)$ defined as:
\begin{equation}
<0|\bar c\gamma_\mu c|V (q)>=f_V(q^2)\epsilon_\mu
\end{equation}
where $\epsilon_\mu$ is the polarization vector of the vector meson $V$.
It has been pointed out recently that in the context of Vector
Meson Dominance, data on photoproduction of $\psi$ indicates a large
suppression of $f_\psi (0)$ compare to $f_\psi (m^2_\psi )$ \cite {dht}.
This has been confirmed independently in Ref \cite {eims} by constraining
the dominant LD contribution to $s\to d\gamma$ using the present upper
bound on the $\Omega^-\to\Xi^-\gamma$ decay rate.  In fact, as we
mentioned earlier, it is argued
that this large suppression results in a much smaller LD contribution
to $b\to s\gamma$ transition.

In the dileptonic rare B-decays, however, $f_V(q^2)$ is normally replaced with
the decay constant $f_V(m^2_V )$ obtained from the leptonic width of
$\psi$ and $\psi'$:
$$
\Gamma (V\to\ell^+\ell^- )=\frac{16\pi\alpha^2}{27m_V^3}f_V^2(m^2_V)
$$
The invariant mass spectrum obtained this way, is dominated by the resonance
interference for a broad range of $q^2$, as already noted in the
literature \cite {osht,amm}.  However, as it was indicated in Ref
\cite{prd}, this spectrum which (using (1)) is written as:
\begin{equation}
\begin{array}{rl}
\displaystyle\frac {1}{\Gamma (B\to X_c e\bar\nu )}\frac {d\Gamma}{dz} (B\to
X_s\ell^+\ell^- ) =& \displaystyle
\left ( \frac {\alpha}{4\pi s_W^2} \right )^2
\frac {2}{f(m_c/m_b)}
\frac {{\vert V_{ts}^*V_{tb}\vert}^2}{{\vert V_{cb}\vert}^2}
{(1-z)}^2 \\ \times &
\displaystyle
\left       ( ( \vert  A \vert^2 +{\vert  B \vert}^2)(1+2z)
                +2\vert  C \vert^2 (1+2/z)\right. \\
     \; &
\displaystyle
\left. +  6Re[{( A + B)}^* C]\right ) \\
\end{array}
\end{equation}
where
$$
f(x)=1-8x^2+8x^6-x^8-24x^4ln(x),
$$
shows a significant reduction in "resonance to nonresonance" interference
when the momentum dependence of $f_V$ (or equivalently, $\psi -\gamma$
transition) is taken into account.  As depicted in fig. 1, at
$q^2/m_b^2\approx 0.3$, for instance, the resonance interference amounts
to around $2\%$ of the differential branching ratio as compared to $20\%$
in the case where fixed $f_V(m^2_V)$ is used.

 In this work, we focus on the parity violating asymmetry distributions
in the inclusive rare decay $B\to X_s\ell^+\ell^-$ when a momentum
dependent $f_V(q^2)
(V=\psi\; ,\; \psi' )$ is inserted in $\tau^{\rm res}$ (we assume
that the same suppression occurs for $\psi'$).  This momentum dependence
is derived in Ref
\cite {terasaki} based on the intermediate quark and antiquark state:
\begin{equation}
f_V(q^2)=f_V(0)\left (1+\frac{q^2}{c_V}\left [d_V -h(q^2)\right ]\right )
\end{equation}
where $c_\psi =0.54\; ,\; c_{\psi'}=0.77$ and $d_\psi =d_{\psi'}=0.043$.
$h(q^2)$ is obtained from a dispersion relation involving the imaginary
part of the quark-loop diagram:
\begin{equation}
\displaystyle
h(q^2)=\frac{1}{16\pi^2r}\left \{
-4-\frac{20r}{3}+4(1+2r)\sqrt{1-\frac{1}{r}}arctan\frac{1}{\sqrt{1-\frac{1}{r}}}\right \}
\end{equation}
with $r=q^2/4m_q^2$ for $0\leq q^2\leq 4m_q^2$.  $m_q$ is the effective
quark mass and assuming that the vector mesons are weakly bound systems of a
quark and an antiquark, we take $m_q\approx m_V/2$.  As a result, eqn (9),
defined for $0\leq q^2\leq m_V^2$, is an interpolation of $f_V$ from the
experimental data on $f_V(0)$(from photoproduction) and $f_V(m_V^2)$(from
leptonic width) based on quark-loop
diagram.  We assume $f_V(q^2)=f_V(m_V^2)$ for $q^2 > m_V^2$ mainly due
to the fact that the behavior of $\psi -\gamma$ conversion strength is
not clear in this region.

The forward-backward asymmetry distribution is defined as:
\begin{equation}
A^{FB}(z)=\frac
{\int_0^1 dw d^2BR/dw dz -\int_{-1}^0 dw d^2BR/dw dz}
{\int_0^1 dw d^2BR/dw dz +\int_{-1}^0 dw d^2BR/dw dz},
\end{equation}
where $w=cos\theta$ with $\theta$ being the angle between the momentum of
the B
meson (or the outgoing s quark) and that of $\ell^+$ in the center of mass
frame of the dileptons.  Using the effective Lagrangian (1) one obtains a
simple form for this asymmetry in the $m_s=0$ limit \cite{amm}:
\begin{equation}
\begin{array}{rl}
A^{FB}(z)=& \displaystyle \frac{3}{2}\frac{
\left       ( ( \vert  A \vert^2 -{\vert  B \vert}^2)z
+  2Re[{( A - B)}^* C]\right )}
{\left       ( ( \vert  A \vert^2 +{\vert  B \vert}^2)(1+2z)
                +2\vert  C \vert^2 (1+2/z)
 +  6Re[{( A + B)}^* C]\right )} \\
=& \displaystyle \frac{3}{2}\frac{
\left (A^{SD}-B^{SD} \right )
\left       ( (  A^{SD} -  B^{SD} + 2Re L)z
+  2 C\right )}
{\left       ( ( \vert  A \vert^2 +{\vert  B \vert}^2)(1+2z)
                +2\vert  C \vert^2 (1+2/z)
 +  6Re[{( A + B)}^* C]\right )} ,
\end{array}
\end{equation}
Where $L=A^{LD}=B^{LD}$.  The forward-backward asymmetry distribution is
shown in fig. 2, where the distributions without $\tau^{res}$ and the one
with constant $\psi$, $\psi' -\gamma$ conversion strength are also
depicted.  We observe that the momentum dependence of the photon-vector
meson coupling results in a small shift in the asymmetry distribution.
In fact, as it is demonstrated, away from the peaks, the exclusion of the
resonances does not change this distribution significantly.

On the other hand, from (12) we observe that $A^{FB}(z)$ is
proportional to $A^{SD}-B^{SD}=A-B$, the coefficient of the leptonic
axial vector current, which does not receive QCD corrections and its
numerical value (2.193 for $m_t=180 GeV$) can be determined accurately
\cite {gsw}.  However, the second factor in the numerator of (12) is
sensitive to QCD corrections through $A^{SD}+B^{SD}+2ReL=Re(A+B)$, and
$C$, the coefficient of the magnetic moment operator.  The latter enters
the rare B-decay $B\to X_s\gamma$.

At this point, we would like to remark that the determination of the
invariant mass $0.1<z_\circ <0.2$, where the forward-backward asymmetry
vanishes, can serve to determine $A^{SD}+B^{SD}\approx -2C/z_\circ$, when
$C$ is known from other channels eg. $B\to X_s\gamma$.  This is due to
the fact
that one can safely ignore the LD effects in this region (see fig. 2).

Next, we turn to the lepton polarization asymmetry distribution which is
defined as:
\begin{equation}
P(z)=\frac
{dBR/dz\vert_{\lambda=-1}-dBR/dz\vert_{\lambda=+1}}
{dBR/dz\vert_{\lambda=-1}+dBR/dz\vert_{\lambda=+1}},
 \end{equation}
with $\lambda =-1$ ($\lambda =+1$) corresponding to the case where the
spin polarization is anti-parallel (parallel) to the direction of the
$\ell^-$ momentum.  In the limit of massless leptons, this simplifies to
left-right asymmetry as left and right-handed leptons do not mix in this
limit.  Therefore, one obtains $dBR/dz\vert_{\lambda =-1}$
($dBR/dz\vert_{\lambda =+1}$) from (8) by setting the coefficients of the
right-handed (left-handed) leptonic current equal to zero.  In this way,
one gets the following expression for the left-right asymmetry:
\begin{equation}
\begin{array}{rl}
A^{LR}(z)=& \displaystyle \frac{
\left       ( ( \vert  A \vert^2 -{\vert  B \vert}^2)(1+2z)
+  6Re[{( A - B)}^* C]\right )}
{\left       ( ( \vert  A \vert^2 +{\vert  B \vert}^2)(1+2z)
                +2\vert  C \vert^2 (1+2/z)
 +  6Re[{( A + B)}^* C]\right )} \\
=& \displaystyle \frac{
\left (A^{SD}-B^{SD} \right )
\left       ( (  A^{SD} -  B^{SD} + 2Re L)(1+2z)
+  6 C\right )}
{\left       ( ( \vert  A \vert^2 +{\vert  B \vert}^2)(1+2z)
                +2\vert  C \vert^2 (1+2/z)
 +  6Re[{( A + B)}^* C]\right )} ,
\end{array}
\end{equation}
We notice that $A^{LR}$, like $A^{FB}$, is proportional to a purely SD
coefficient ie., $A^{SD}-B^{SD}$.  The
asymmetry $A^{LR}(z)$ is shown in fig. 3 for cases where i) momentum
dependence of $f_V(q^2)$ is taken into account, ii) $f_V(q^2)$ is
replaced with fixed $f_V(m_V^2)$ and iii) $\tau^{res}$ is excluded.  We
observe that the asymmetry distribution, away from the peaks, receives only
small interference (the deviation from the curve with $\tau^{res}$
set to zero) from the resonances.  For example, at $z=q^2/m_b^2=0.3$,
this interference amounts only to 2\% of the asymmetry at this invariant
mass.

In conclusion, we investigated the effect of a momentum dependent
photon-vector meson coupling on various distributions of the dileptonic
rare B-decay $B\to X_s\ell^+\ell^-$.  The effect is more significant in
the invariant mass spectrum (fig. 1),
resulting only in a small shift in the forward-backward (fig. 2) and
left-right (fig. 3)
asymmetry distributions.  In light of these results, we believe that one
can get reliable complementary information about SD physics from these
distributions.

\vskip 2.0cm
{\bf \Large Acknowledgement}
\vskip 0.5cm
The author would like to thank R. R. Mendel for useful discussions.

\newpage
\begin{flushleft}
\bf \Huge Figure Caption
\end{flushleft}
{\bf \Large Figure 1:}
The dileptonic invariant mass spectrum for the decay
$b\to
s\ell^+\ell^-$.  The thin, dotted and bold lines correspond to the spectrum
without resonances, with resonances but constant $V-\gamma$ conversion
strength and with resonances having momentum dependent $V-\gamma$
transition respectively.  For $q^2 >m_{\psi'}^2$, where the latter two
curves coincide, only the dotted curve is shown.\\

{\bf \Large Figure 2:}
The forward-backward asymmetry distribution for the decay
$b\to
s\ell^+\ell^-$.  The thin, dotted and bold lines correspond to the asymmetry
without resonances, with resonances but constant $V-\gamma$ conversion
strength and with resonances having momentum dependent $V-\gamma$
transition respectively.  For larger values of $z$, where the latter two
curves coincide, only the dotted curve is shown.\\

{\bf \Large Figure 3:}
The polarization (left-right) asymmetry distribution for the decay
$b\to
s\ell^+\ell^-$.  The thin, dotted and bold lines correspond to the asymmetry
without resonances, with resonances but constant $V-\gamma$ conversion
strength and with resonances having momentum dependent $V-\gamma$
transition respectively.  For larger values of $z$, where the latter two
curves coincide, only the dotted curve is shown.

\newpage
\setlength{\unitlength}{0.240900pt}
\ifx\plotpoint\undefined\newsavebox{\plotpoint}\fi
\sbox{\plotpoint}{\rule[-0.200pt]{0.400pt}{0.400pt}}%


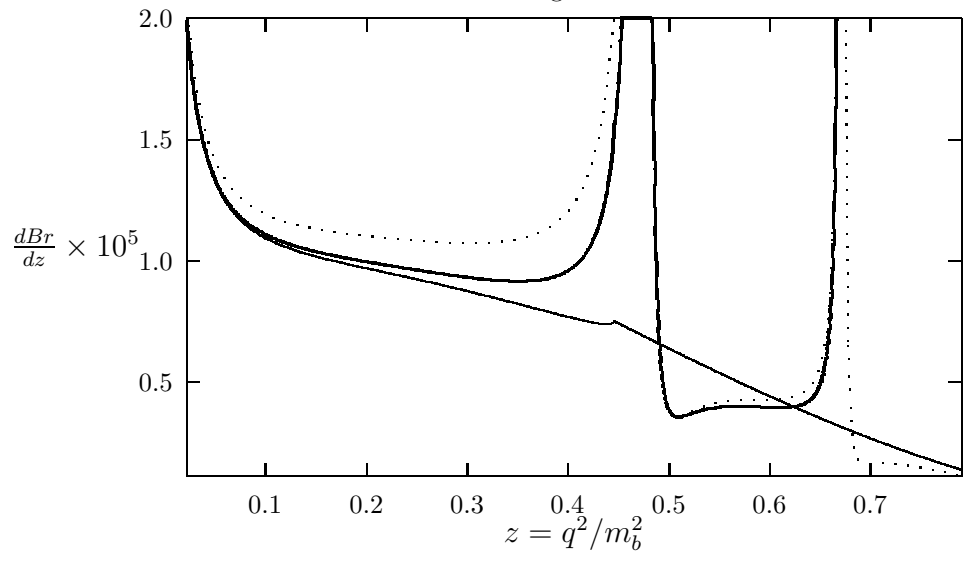
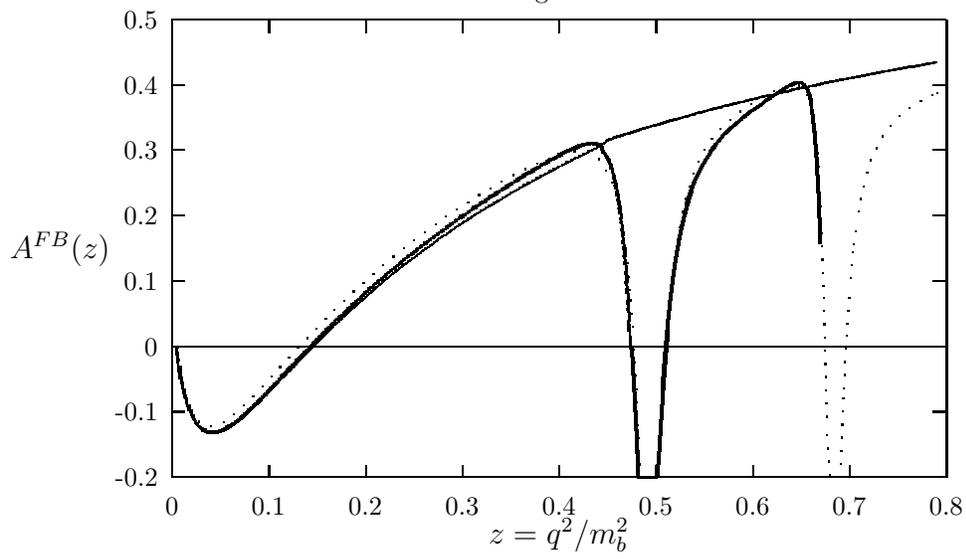
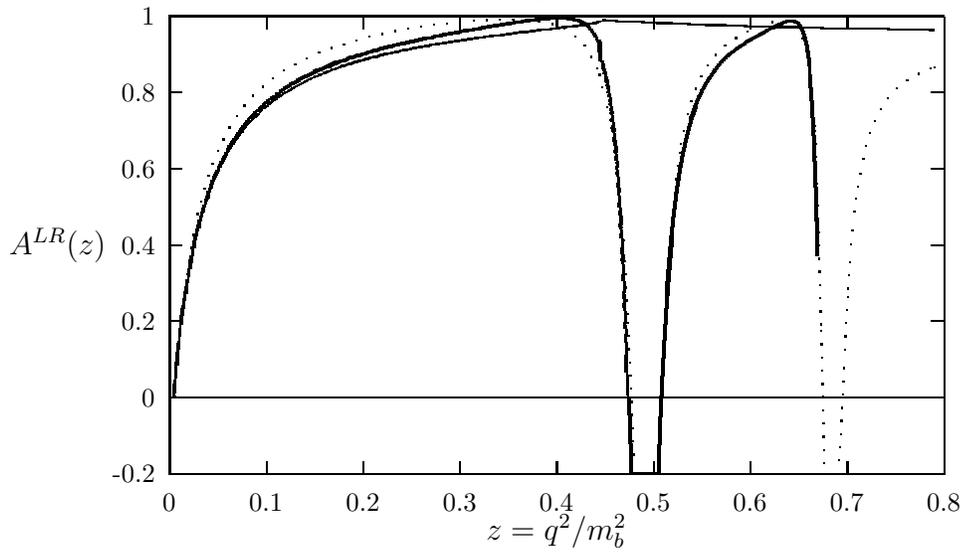
\end{document}